# Prediction of High Temperature Quantum Anomalous Hall Effect in Two Dimensional Transition-Metal Oxides

H. P. Wang[1,3], Wei Luo[1,2,3], and H. J. Xiang[1,2,3*]

[1]*State Key Laboratory of Surface Physics, and Department of Physics, Fudan University, Shanghai 200433, P. R. China*

[2]*Key Laboratory of Computational Physical Sciences (Ministry of Education), Fudan University, Shanghai 200433, P. R. China*

[3]*Collaborative Innovation Center of Advanced Microstructures, Nanjing 210093, P. R. China*

Email: hxiang@fudan.edu.cn

**Abstract**

Quantum anomalous Hall (QAH) insulator is a topological phase which exhibits chiral edge states in the absence of magnetic field. The celebrated Haldane model is the first example of QAH effect, but difficult to realize. Here, we predict the two-dimensional single-atomic-layer $V_2O_3$ with a honeycomb-Kagome structure is a QAH insulator with a large band gap (large than 0.1 eV) and a high ferromagnetic Curie temperature (about 900 K). Combining the first-principle calculations with the effective Hamiltonian analysis, we find that the spin-majority $d_{xy}$ and $d_{yz}$ orbitals of V atoms on the honeycomb lattice form a massless Dirac cone near the Fermi level which becomes massive when the on-site spin-orbit coupling is included. Interestingly, we find that the large band gap is caused by a cooperative effect of electron correlation and spin-orbit coupling. Both first-principle calculations and the effective Hamiltonian analysis confirm that 2D $V_2O_3$ has a non-zero Chern number (i.e., one). Our work paves a new direction towards realizing the QAH effect at room temperature.

Quantum anomalous Hall (QAH) effect is a quantized version of anomalous Hall effect (AHE) [1], in which the Hall conductance is quantized in the form of $ve^2/h$ ($v$ is an integer) even in the absence of an external magnetic field. Different from the traditional quantum Hall effects from Landau-level quantization [2], the QAH effect (QAHE) is a consequence of the combination of spontaneous magnetization and spin-orbit coupling (SOC). Because of the presence of the chiral edge state that can carry electric current without dissipation and spin polarization in a QAH system [3], the realization of QAHE may lead to the development of a new generation of low-power-consumption electronics and novel spintronic devices.

For a long time, the quantum Hall effect was thought to exclusively emerge in the presence of an external magnetic field. In 1988, Haldane proposed the possibility of QAHE by considering a honeycomb lattice with a staggered magnetic-flux [4]. Unfortunately, due to the special requirements, Haldane's model is difficulty to realize in real materials although it was realized recently with ultracold fermionic atoms [5]. Hence, it is of great interest to discover the QAH state in realistic materials. As we know, to be a QAH insulator (i.e., Chern insulator), it must satisfy four conditions: two-dimensional (2D), insulating in the bulk, broken time-reversal symmetry (TRS) with a certain magnetic ordering [e.g., a ferromagnetic (FM) order)], a non-zero Chern number for the occupied bands. It is not difficult for a material to satisfy one or two conditions. But finding a realistic material that satisfies all of these conditions simultaneously turns out to be very challenging. The discovery of quantum spin Hall effect (QSHE) and topological insulators (TIs) [6-11] have greatly boosted the development of field of QAHE. Several realistic systems, including Mn-doped HgTe quantum wells [12] and magnetic-impurity doped $Bi_2Se_3$ thin films [13], have been proposed to be QAH insulators. The essence of these proposals is to break the TRS in a 2D slab of TIs by magnetic doping. If the FM order can be established, the introduced exchange splitting may destroy the band inversion for one spin channel, while keeping the band inversion in the other spin channel, leading to the QAHE. Subsequently, Chang *et al.* [14] observed the quantization of the Hall resistance at $h/e^2$ at zero external magnetic field with temperature around 30 mK in Cr-doped $(Bi,Sb)_2Te_3$ thin films. However, the extremely low temperature hinders realistic applications. Therefore, searching for Chern insulators with high Curie temperature and large band gap is essential to realize the high temperature QAHE. Some theoretical proposals were suggested to realize the high-temperature QAHE systems [15-21]. In particular, Garrity and Vanderbilt demonstrated that the GaN/EuO interface is a candidate for achieving a QAH state at temperature of around 70 K and with band gaps of about 0.1 eV [22]. Wu *et al.* proposed that half-passivated stanene and germanene become Chern insulators with energy gaps of 0.34 and 0.06 eV, respectively [23]. The proposed FM order in this system comes from the unpassivated sublattice that exhibits dangling *p* orbitals. After applying a correction on the mean-field results, the Curie temperatures were estimated to be 148 and 310 K for stanene and germanene, respectively. Recently, QAHE was also predicted in antiferromagnetic (AFM) systems such as 2D $NiRuCl_6$ [24] and $Sr_2FeOsO_6$ films [25]. QAH insulators with AFM coupling might be important for the search of new topological superconducting materials and might have high magnetic transition temperature. Despite of these theoretical progress, more theoretical and experimental efforts are certainly required for finding the route toward a room temperature QAHE.

In this work, we predict that 2D $V_2O_3$ with the honeycomb-Kagome (HK) lattice is a room temperature QAH insulator. Our Monte Carlo simulation shows that 2D $V_2O_3$ has a FM Curie temperature up to 900 K. The non-trivial band gap exceeds 0.1 eV, greater than the room temperature energy scale ($k_BT \approx 26$ meV at T = 300 K). The band gap opening is found to be a cooperative effect of electron correlation and spin-orbit coupling. First-principle calculations of the Berry curvature confirm that the FM state of 2D $V_2O_3$ has a Chern number of 1. Phonon and molecular dynamics (MD) simulations show that 2D $V_2O_3$ is dynamically and kinetically stable. A tight-binding model and low-energy effective Hamiltonian are constructed to understand the unusual electronic properties. As an unusual oxide Chern insulator, 2D $V_2O_3$ is naturally stable against oxidization and degradation. Given all these excellent properties of 2D $V_2O_3$, we expect that it will be a promising candidate for realizing room temperature QAHE.

The proposed 2D $V_2O_3$ adopts the HK lattice with the P6m plane group, as shown in Fig. 1(a). The HK structure [26] was predicted to host a 2D node-line semimetal state in $Hg_3As_2$. The HK $A_2B_3$ lattice with five atoms per unit cell can be obtained by inserting B atoms in between two neighboring A atoms of the honeycomb A lattice. Therefore, 2D $V_2O_3$ has the same $D_{6h}$ point group symmetry as graphene. Interestingly, 2D planar $Al_2O_3$ monolayer with this HK lattice was found to be highly stable [27].

Based on the first-principle calculations, we calculate the phonon spectrum to confirm the structure stability. There are no imaginary frequency phonons in the whole Brillouin zone, which means that this structure is dynamically stable (see Supplemental Material [28]). Meanwhile, *ab initio* MD simulations are carried out in a $5 \times 5 \times 1$ supercell at the 300 K to verify its thermal stability. For 3000 time steps with a time step of 1.5 fs, the total energies are almost unchanged during the simulation (see Supplemental Material [28]) and the structure only has the slight distortions after the simulation (see Supplemental Material [28]), suggesting that its structure is thermally stable. These results indicate that 2D $V_2O_3$ is stable at room temperature.

Next, we focus on the magnetic properties of the $V_2O_3$ monolayer. Our DFT calculations show that the ground state of $V_2O_3$ is FM. The FM state has a lower energy than the paramagnetic nonmagnetic state and antiferromagnetic Neel state [see Fig. 1(a)] by 0.79 eV/V and 0.53 eV/V, respectively. The density of states (DOS) of the FM ground state for $V_2O_3$ from the LDA+U calculation is shown in Fig. 2(a). It can be seen that the states near the Fermi level are dominated by V 3d states. The O 2p states locate between -7.5 eV and -3 eV. At the LDA+U level, $V_2O_3$ is a half metal, i.e., it is metallic for the spin-up channel, but insulating for the spin-down channel. Detailed local DOS analysis shows that the $V^{3+}$ ions take the high-spin electronic configuration as shown in Fig. 2(b). The energies of the orbitals increase in the order of $d_{z2}$, ($d_{xz}$, $d_{yz}$), ($d_{x2-y2}$, $d_{xy}$). Note that $d_{xz}$ and $d_{yz}$ ($d_{x2-y2}$ and $d_{xy}$) orbitals are degenerate due to the three-fold symmetry of V-site. This energy order can be explained by considering the orbital interactions between O 2p orbitals and V 3d orbitals. Note that if the interaction between the low-lying O 2p orbitals and a 3d orbital is stronger, the 3d orbital is pushed to a higher energy level. The V $d_{z2}$ orbital has the lowest energy since the interaction between the O 2p state and this d orbital in a planar structure is negligible. The V ($d_{xz}$, $d_{yz}$) orbitals form π bonds with O $2p_z$ orbitals, while the V ($d_{x2-y2}$, $d_{xy}$) orbitals form σ bonds with O ($p_x$, $p_y$) orbitals. The fact that π interaction is weaker than σ interaction explains why ($d_{x2-y2}$, $d_{xy}$) orbitals has the highest energy. The high-spin $V^{3+}$ ion has two valence electrons, resulting a spin magnetic moment of 2 $\mu_B$. One electron occupies the spin-up $d_{z2}$ orbital, and the other one fills half of the spin-up ($d_{xz}$, $d_{yz}$) level. Below we will show that the peculiar electronic configuration leads to a strong ferromagnetism in $V_2O_3$.

The extremely strong ferromagnetism is because the spin-majority ($d_{xz}$, $d_{yz}$) levels are half filled. As can be seen in Fig. 2(c), the spin-up ($d_{xz}$, $d_{yz}$) orbital of a V ion ($V_1$) can interact with the spin-up ($d_{xz}$, $d_{yz}$) orbital of a neighboring V ion ($V_2$) through the shared O atom in Fig.1(a). This interaction will lead to a two-fold degenerate low-lying bonding state and another two-fold degenerate high-lying anti-bonding state. The two electrons will occupy the two-fold degenerate bonding state, resulting in an energy lowering [ $\Delta$ E(FM)∝t, where t is the effective hopping between the ($d_{xz}$, $d_{yz}$) orbitals of $V_1$ and $V_2$] when the two V ions have the same FM spin alignment. This is similar to the well-known double exchange mechanism for ferromagnetism. When the spins of the two V ions are antiparallel to each other, there is an usual energy lowering [ $\Delta$ E(AFM)∝$t^2$/U] due to the superexchange mechanism. Since U is much larger than t, $\Delta$ E(FM) is much larger than $\Delta$ E(AFM). This explanation is further confirmed by the first-principle calculations. We find that the energy difference between the Neel AFM state and the FM state increases with the Hubbard U value [see Fig. 3a]. This is because the energy lowering in the AFM case decreases with U, while that in the FM case does not depend on U. The stability of the FM state is also robust with respect to the exchange J parameter of the LDA+U method (see Fig. S3 of Supplemental Material [28]).

We include spin-orbit coupling in the LDA+U calculation to examine the magnetic anisotropy of $V_2O_3$. It turns out that $V_2O_3$ has a strong easy-axis magnetic anisotropy with the spins along the z axis [see Fig. 3b]. The FM state with the spins along z has a lower energy by 2.0 meV/f.u. than that with the spins along the in-plane direction. The magnetic anisotropy increases when the Hubbard U value becomes larger since the orbital moment becomes larger [see Fig.3(c)]. The easy-axis anisotropy can be understood from the degenerate perturbation theory. Since $\langle d_{xz}|\hat{L}_z|d_{xy}\rangle = -i$, while $\langle d_{xz}|\hat{L}_+|d_{xy}\rangle = \langle d_{xz}|\hat{L}_-|d_{xy}\rangle = 0$, the spin-conserving term [29] of spin-orbit coupling operator $\hat{S}_n(\hat{L}_z\cos\theta + \frac{1}{2}\hat{L}_+e^{-i\phi}\sin\theta + \frac{1}{2}\hat{L}_-e^{i\phi}\sin\theta)$ [$\theta$ and $\phi$ are the zenith and azimuth angles of the magnetization in the direction $n(\theta,\phi)$] can lower the energy when all the spins are along the z-axis. The easy-axis anisotropy is important to the presence of QAH state in 2D $V_2O_3$ since the band gap remains zero in the case of in-plane magnetization, as will be discussed later.

From the energy difference between the FM state and AFM Neel state [see Fig. 1(a)], we can estimate the effective spin exchange parameter for the nearest neighboring V-V pair to be J = -0.17 eV. The appropriate spin Hamiltonian for $V_2O_3$ is $H = \sum_{<NN\,i,j>} J\vec{S}_i \cdot \vec{S}_j + \sum_i AS_{iz}^2$, where the single-ion anisotropy parameter A is -2 meV. With this spin Hamiltonian, we performed parallel tempering Monte Carlo simulation [30,31] to estimate the Curie temperature of $V_2O_3$ to be 900K [see Fig.1(b)]. Therefore, we predict that 2D $V_2O_3$ is a room-temperature ferromagnet.

Now let us focus on the electronic properties of the FM state of 2D $V_2O_3$. Fig. 4(a) shows the LDA+U band structure. When SOC is not taken into account, 2D $V_2O_3$ is a half-metal. Interestingly, there is a Dirac cone at K (also at $K' = -K$), similar to the graphene case. Different from the graphene case where the well-known Dirac dispersion is due to the $p_z$ orbital, here the crossing bands are mainly of the ($d_{xz},d_{yz}$) characters. Note that there are four spin-up ($d_{xz},d_{yz}$) bands with the other two bands almost flat. We will show that the peculiar band dispersion is a consequence of the interaction between the ($d_{xz},d_{yz}$) orbitals in the honeycomb lattice, similar to the case of ($p_x,p_y$) orbitals in the honeycomb lattice [32]. When SOC effect is included in the calculation, the degeneracy at the Dirac points is lifted [see Figs. 4(b) and 4(c)] when the magnetization is along the z-axis. With U = 4 eV, 2D $V_2O_3$ is an indirect-gap insulator with the VBM and CBM at K and Γ, respectively. The direct and indirect gaps are found to be 0.22 eV and 0.12 eV, respectively. To the best of knowledge, 2D $V_2O_3$ is one of the rare QAH insulators with a band gap larger than 0.1 eV [22-24, 33, 34].

We note that the band gap of 2D $V_2O_3$ depends on the Hubbard U value: Stronger electron correlation (i.e., larger U) leads to a larger band gap [see Fig. 3(d)]. Due to the poor screening of the Coulomb interaction in this single-layer system, it is expected that the Hubbard U in 2D $V_2O_3$ would be larger than 4 eV, thus its real band gap should be larger than 0.1 eV. When the SOC is switch off, the system still keeps the Dirac-cone even when the U value is as high as 6 eV. The band gap opening is a result of the cooperative effect of electron correlation and SOC [compare Fig. 4(b) and 4(c)], which was first found in $Ba_2NaOsO_6$ [35] and later in $Sr_2IrO_4$ [36] and bilayer system of $LaCoO_3$ [37]. First, the degeneracy is split by SOC, then the splitting is further enhanced by the electron correlation [see Fig. 4(d)]. The band gap becomes even larger if a smaller exchange J parameter of the LDA+U method is adopted (see Fig. S3 of Supplemental Material [28]).

The Dirac-type dispersion in the absence of SOC and the band gap opening by SOC suggests that 2D FM $V_2O_3$ might be a Chern insulator. In order to verify this, we construct the MLWF and calculate the Berry curvature. By integrating the Berry curvature in the whole Brillouin zone, we find that the Chern number C = 1. These first-principle calculations confirm that 2D FM $V_2O_3$ is indeed a Chern insulator. Therefore, our above results show that 2D $V_2O_3$ is a room-temperature Chern insulator.

In the literature, a common mechanism which gives rise to topological bulk band gaps is the band inversion between the cation state and anion state [38, 39]. However, the nontrivial band topology in $V_2O_3$ monolayer results from the massless Dirac cone rather than the usual band inversion. In fact, the origin of the nontrivial topology in $V_2O_3$ monolayers resembles that of graphene and silicone [40]. However, there is an on-site SOC in 2D $V_2O_3$, which is different from the high-order effect as in graphene and silicene. In order to unveil the origin of the topological electronic structure of 2D $V_2O_3$, we carry out detailed tight-binding (TB) and effective Hamiltonian analysis. Based on the first-principle calculations, the low-energy band structure mainly comes from V $d_{xz}$ and V $d_{yz}$ orbitals in the spin up channels. The following TB model $H = H_0 + H_{so}$ can be constructed to explain the band structure by considering spin-up

V $d_{xz}$ and V $d_{yz}$ orbitals. At the level of the nearest neighbor hopping approximation, the hopping part of the Hamiltonian is

$$H_0 = t\sum_{\vec{r}\in A}\{d^{\dagger}_{i,\uparrow}(\vec{r})d_{i,\uparrow}(\vec{r}+a\vec{e}_i)+h.c.\}, \tag{1}$$

where $\vec{e}_{1,2} = \pm\frac{\sqrt{3}}{2}\vec{e}_x + \frac{1}{2}\vec{e}_y$ and $\vec{e}_3 = -\vec{e}_y$ are three unit vectors from A site to its three neighboring B sites [Note that there are two distinct V sites, A and B, in the unit cell, see Fig. 1(a)]; $a$ represents the distance between neighboring A and B sites; $d_i = (d_{xz}\vec{e}_x + d_{yz}\vec{e}_y)\cdot\vec{e}_i (i=1,2,3)$ are the projections of d-orbitals along the $\vec{e}_i$ direction. In the basis we choose, the on-site SOC $H_{so} = -\lambda\vec{S}\cdot\vec{L}$ ($\lambda > 0$) reads

$$H_{so} = i\lambda\sum_{\vec{r}}\{d^{\dagger}_{xz,\uparrow}(\vec{r})d_{yz,\uparrow}(\vec{r}) - d^{\dagger}_{yz,\uparrow}(\vec{r})d_{xz,\uparrow}(\vec{r})\}. \tag{2}$$

Due to the particle-hole symmetry of the Hamiltonian, the band structure is symmetric with respect to the zero energy. In the absence of SOC, there are both flat bands and Dirac cones in the band dispersion [see Fig. 5(a)]. The flat bands and dispersive bands touch at the center of the first BZ, and two dispersive bands touch at Dirac cones. When the SOC is turned on, the band gap opens. The degeneracy between the first and second bands is lifted. The lowest band is no longer flat and the second band is pushed up as shown in Fig. 5(c). The degeneracy between the middle two dispersive bands at the Dirac cones is also lifted. It is clear that along the k-path $M\rightarrow\Gamma\rightarrow K\rightarrow M$, there is a good agreement with the first-principle results. To gain more insight into the origin of topological nature near the $K$ and $K'$ points, a low energy effective Hamiltonian (LEEH) is derived for the middle two bands near the Fermi level. With the two bases $\Phi_1 = d_{A,-\uparrow}$ and $\Phi_2 = d_{B,+\uparrow}$ ($d^+_{\pm\uparrow} = (d^+_{xz\uparrow}\pm id^+_{yz\uparrow})/\sqrt{2}$), we obtain

$$H^K_{eff}(\vec{k}) = \sum_i b_i(\vec{k})\sigma_i \tag{3}$$

where $\sigma_{i=1,2,3}$ are Pauli matrices and $b_1 = 3k_x t/4, b_2 = 3k_y t/4, b_3 = \lambda$. This is similar to Dirac Hamiltonian in graphene. The eigenvalues of Hamiltonian is $E(k) = \pm\sqrt{\lambda^2 + 9t^2k^2/16}$, which indicates that the system is gapless and the dispersive relation is linear in the absence of the SOC [see Fig. 5(b)], and a band gap will be opened at the $K$ point when SOC is included [see Fig. 5(d)]. Similarly, we can obtain the LEEH near the $K'$ point $H_{eff}^{K'}(\vec{k}) = \sum_i b_i(\vec{k})\sigma_i$ with $b_1 = -3k_x t/4, b_2 = 3k_y t/4, b_3 = -\lambda$. To characterize the band topology, we calculate Berry curvature near the $K$ and $K'$ points [41, 42]. By integrating the Berry curvature, we find that the Chern number is 1 (see Supplemental Material [28]), consistent with the first-principle calculation. We note that the low-energy physics in 2D $V_2O_3$ is similar to the case of $p_x$ and $p_y$ orbitals on the honeycomb lattice, which was found to lead to new phenomena, such as the quantum anomalous Hall effect [43] and Wigner crystallization [32]. However, usually the $p_x,p_y$ orbitals could not lead to a robust FM order.

In summary, based on the first-principle calculations, we have systematically investigated the stability, magnetic and electronic structures of the 2D HK $V_2O_3$. The phonon calculations and ab initio MD simulations confirm that the $V_2O_3$ monolayer is dynamically and thermally stable. The Curie temperature is estimated to be 900K by using the Monte Carlo simulations. We predict that it is a robust intrinsic FM QAH insulator with a bulk band gap larger than 0.1 eV. Their topological characteristic is confirmed by computing the Chern number with first-principle method and the low energy effective model. Interestingly, the large band gap in the $V_2O_3$ is not caused by the large SOC magnitude itself, but by a cooperative effect of electron correlation and SOC. Besides its robust topological electronic structures, 2D $V_2O_3$ is naturally stable against oxidization and degradation, cheap, non-toxic without heavy elements. Our work suggests that 2D 3d transition-metal oxides are promising candidates for high temperature QAH insulators.

This work was supported by NSFC (11374056), the Special Funds for Major State Basic Research (2012CB921400, 2015CB921700), Program for Professor of Special Appointment (Eastern Scholar), Qing Nian Ba Jian Program, and Fok Ying Tung Education Foundation. We thank Prof. R. B. Tao for useful discussion.

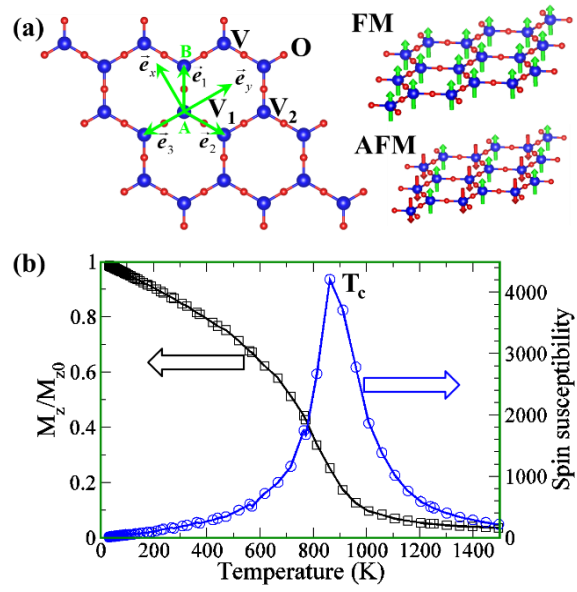

FIG. 1. (a) Lattice structure for the 2D $V_2O_3$ monolayer and schematic representations of FM and Neel AFM orders. (b) The magnetic moment (black) and spin susceptibility (blue) as a function of temperature from the Monte Carlo simulation, indicating that the ferromagnetic transition temperature is very high (about 900 K).

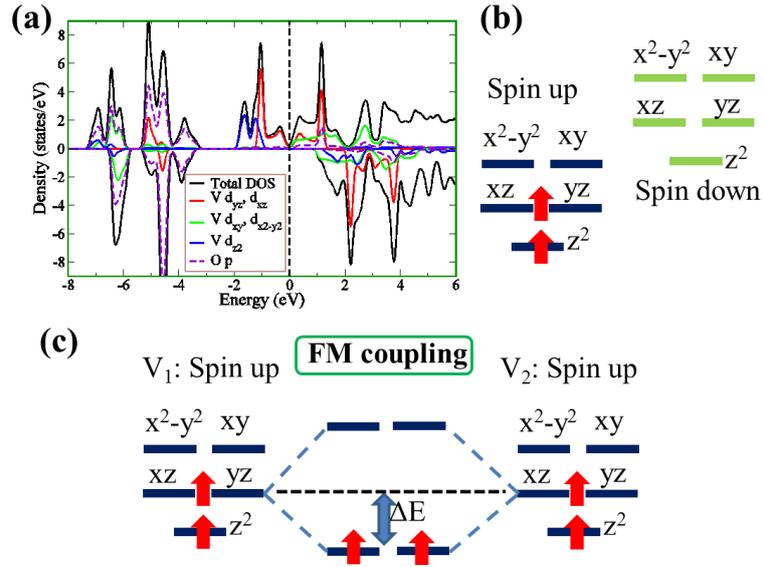

FIG. 2. (a) DOS and PDOS of $V_2O_3$ calculated at the LDA+U level. From it, one can easily see the bands near the Fermi level are dominated by $d_{xz}$ and $d_{yz}$ orbitals. (b) Electronic configuration of $V^{3+}$ in $V_2O_3$. The spin down 3d states is far away from the Fermi level because of the exchange splitting. (c) The schematic illustration of exchange mechanism for the FM coupling. It is clear that the occupation of the low-lying bonding states lowers the total energy.

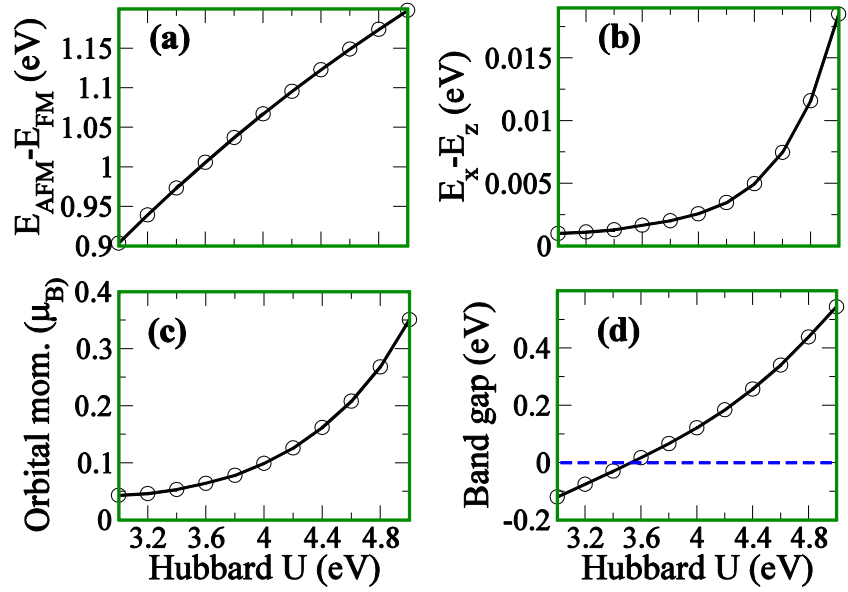

FIG. 3. Dependence of physical properties on the Hubbard U value. (a) Energy difference (eV/f.u.) between AFM and FM states. (b) Energy difference between the FM state with spins along $x$-axis and that with spins along $z$-axis, indicating that 2D $V_2O_3$ has an easy-magnetization axis along $z$. (c) Local orbital moments of $V^{3+}$ ion. (d) The global band gap. Note that there is no band gap when the spins are along the in-plane directions.

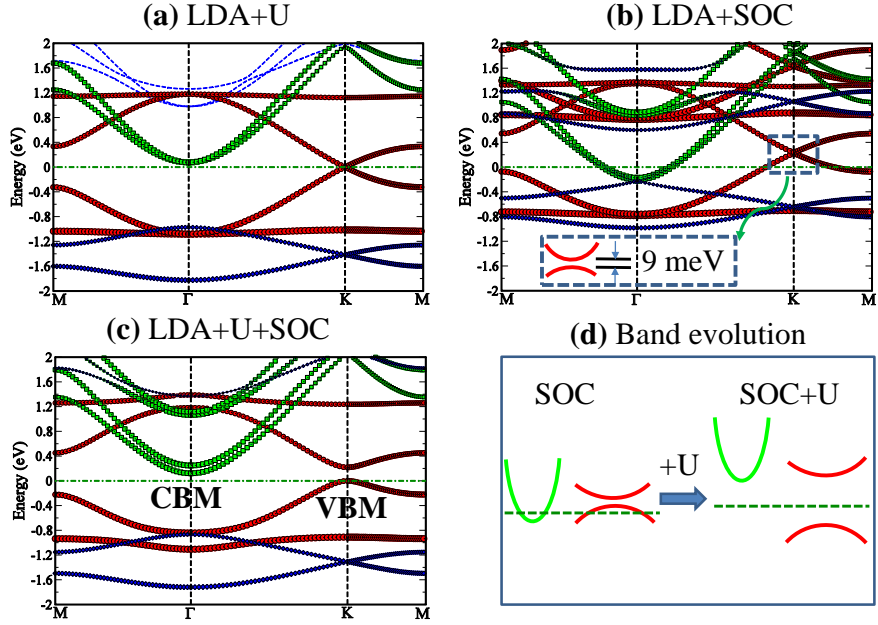

FIG. 4. Band structures of 2D $V_2O_3$ calculated using different methods. (a) LDA+U with U = 4 eV, (b) LDA+SOC and (c) LDA+U+SOC with U = 4eV. The bands near the Fermi level consist of the V 3d orbitals and the Fermi level is set to zero. (d) Schematic illustration of the band evolution by considering SOC and Hubbard U interaction.

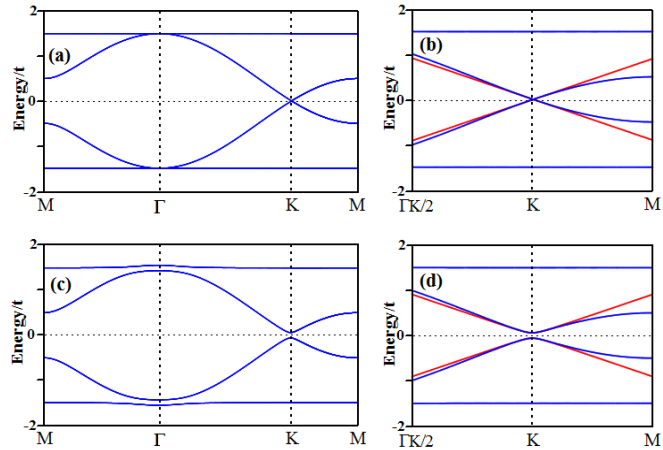

FIG. 5. Band structures of 2D $V_2O_3$ calculated using the TB model. (a) and (b): without SOC. (c) and (d): with SOC. The band structures near the K point from the LEEH (solid red) are also shown in (b) and (d). The horizontal dotted lines indicate the Fermi level. The band structure from the TB model agrees with the first principles results [see red circles in Figs. 4(a) and 4(c)]. One can see the results near K from the TB model and LEEH are in agreement with each other very well.